\documentclass{article}

\usepackage{microtype}
\usepackage{graphicx}
\usepackage{booktabs} 
\usepackage{amsmath}
\usepackage{multirow}
\usepackage{stfloats}
\usepackage{subcaption}
\usepackage{graphicx}

\usepackage{hyperref}

\usepackage[accepted]{mlsys2024}

\mlsystitlerunning{HeteGen: Heterogeneous Parallel Inference for Large Language Models on Resource-Constrained Devices}

\begin{document}

\twocolumn[
\mlsystitle{HeteGen: Heterogeneous Parallel Inference for Large Language Models on Resource-Constrained Devices}



\mlsyssetsymbol{equal}{*}

\begin{mlsysauthorlist}
\mlsysauthor{Xuanlei Zhao}{equal,nus}
\mlsysauthor{Bin Jia}{equal,nus}
\mlsysauthor{Haotian Zhou}{equal,nus}
\mlsysauthor{Ziming Liu}{nus}
\mlsysauthor{Shenggan Cheng}{nus}
\mlsysauthor{Yang You}{nus}
\end{mlsysauthorlist}

\mlsysaffiliation{nus}{National University of Singapore}

\mlsyscorrespondingauthor{Yang You}{youy@comp.nus.edu.sg
}

\mlsyskeywords{Machine Learning, MLSys}

\vskip 0.3in

\begin{abstract}
  In recent times, the emergence of Large Language Models (LLMs) has resulted in increasingly larger model size, posing challenges for inference on low-resource devices. Prior approaches have explored offloading to facilitate low-memory inference but often suffer from efficiency due to I/O bottlenecks. To achieve low-latency LLMs inference on resource-constrained devices, we introduce HeteGen, a novel approach that presents a principled framework for heterogeneous parallel computing using CPUs and GPUs. Based on this framework, HeteGen further employs heterogeneous parallel computing and asynchronous overlap for LLMs to mitigate I/O bottlenecks. Our experiments demonstrate a substantial improvement in inference speed, surpassing state-of-the-art methods by over 317\% at most.
\end{abstract}
]

\printAffiliationsAndNotice{\mlsysEqualContribution} 
\section{Introduction}
In recent years, Large Language Models (LLMs) have exhibited remarkable performance improvements, correlating with the exponential growth in their scale \cite{gpt3, bert_2019, opt_2022, instructgpt}. These models have not only enhanced their fundamental capabilities but have also demonstrated the emergence of novel functionalities \cite{emergent}, rendering them more proficient across a wide array of tasks. The substantial grow in model size has introduced considerable memory demands for their deployment. The deployment of most LLMs now necessitates tens, or even hundreds, of gigabytes of memory for inference, leading to the biggest barrier for their application.


In this paper, our focus is to achieve low-latency inference for LLMs on resource-constrained devices. In many applications, we requires the model to be deployed on local device for reasons like stability, privacy, safety and systems constraints. These situations include individual user access, autonomous driving \cite{drive_2023}, edge computing \cite{edgeai_2019}, and personal assistant \cite{ross_programmers_2023}. The key characteristics of such tasks are as follows: 1) Their GPU resources are often limited on local hardware. 2) They typically use small batch sizes, typically 1, but require quick responses for an enhanced user experience, which presents a challenge for existing techniques.

\begin{figure}[t]
    \centering
    \includegraphics[width=0.7\linewidth]{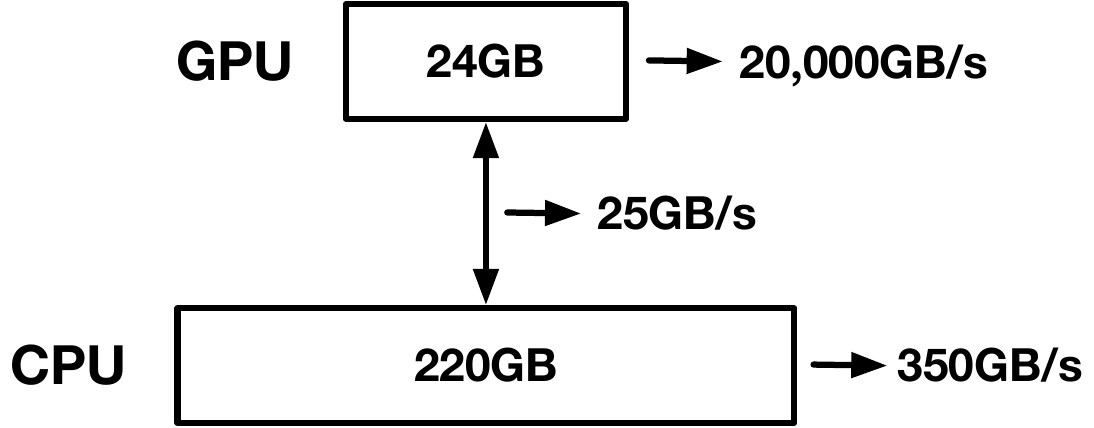}
    \caption{Memory space and processing speed for GPU, CPU, and I/O between CPU and GPU. The speed is tested with OPT-30B MLP Linear on NVIDIA A10 GPU and Intel Xeon @ 2.30GHz CPU, calculated as parameter size divided by processing time.}
    \label{fig:speed}
\end{figure}

Model compression \cite{llmint8_2022, smoothquant, gptq} compress the parameter weight's bitwidth to reduce parameter memory, and low-level optimization including KV-cache management \cite{vllm_2023} and fused attention kernels \cite{flashattention_2022} have been proposed to reduce memory usage by system optimization. But their capability to save memory is still far from expected. ZeRO-offload \cite{zero-offload, deepspeed_inference} proposes to offload unused parameters to CPU memory and disk to reduce memory cost significantly, but it leads noticeable loss in speed. FlexGen \cite{flexgen} improves offloading throughput for large batch inference by computing attention in CPU and overlapping I/O with computation. However, FlexGen's utilization of CPU and I/O resources is still limited, and it doesn't efficiently reduce latency for sparse inputs.

As illustrated in Figure \ref{fig:speed}, the CPU's memory capacity significantly surpasses that of the GPU, allowing it to accommodate unused parameters for the GPU. Additionally, the I/O between the GPU and CPU is considerably slower compared to the GPU's computational efficiency, thus acting as a bottleneck within the system. However, in such a system, the potential of the CPU is often underestimated, resulting in either idle or very light workload. Our key observation is that since the bottleneck of LLM inference with offloading is I/O, we can leverage the CPU for heterogeneous parallel computation alongside the GPU, thereby reducing the need for parameter I/O and achieving an better resource allocation. In this context, there are three core challenges that need to be addressed: 1) What parallel strategy should be employed for CPU and GPU? 2) How to distribute computation between the CPU and GPU to achieve optimal efficiency? 3) How to improve the usage of I/O and CPU computation?

To solve these challenges, we propose HeteGen, a heterogeneous parallel inference system that can effectively reduce the latency of LLMs on low-resource devices by fully utilizing CPU and I/O resources. HeteGen proposes a general framework for heterogeneous parallel computing using both CPUs and GPUs. Building upon this framework, HeteGen incorporates heterogeneous parallel computing and asynchronous overlap to address I/O bottlenecks in Large Language Models (LLMs) inference. Our experiments show a significant enhancement in inference speed, exceeding state-of-the-art methods by  317\%.

In summary, our contributions are as follows:
\begin{itemize}
\item We proposed a low-latency offloading inference approach based on heterogeneous parallel computing asynchronous overlapping to alleviate the bottleneck of I/O.
\item We have developed a general heterogeneous parallel method and corresponding theoretical formula that guides optimal performance in computations distribution for heterogeneous parallel computing.
\item HeteGen incorporates heterogeneous parallel computing and asynchronous overlap to address I/O bottlenecks in Large Language Models (LLMs) inference.
\item Our experimental results demonstrate that we have surpassed the current state-of-the-art methods in latency and dynamic ranges.
\end{itemize}

\section{Background}
\subsection{Generative Language Models Inference}

In this section, we present the inference process of generative language models. The Transformer \cite{attention} takes an input sequence, which undergoes a multi-head self-attention mechanism to produce a context-sensitive representation. This representation then passes through forward network layers, repeating the above steps until the final prediction is generated. During multi-head attention, we compute the query, key, and value based on our input and calculate the attention result according to a specific formula. After attention, two feed-forward networks (FFN) further process the features in MLP module. 

We introduce the two stages of inference for generative language models, prefill and decode, and discuss their differences. In the prefill stage, the generative model receives the input text (also known as "prompt") and processes it in parallel. Our input is $[x_1, x_2, ... x_n]$, which is processed according to the Transformer's standard inference flow. At each attention layer, we save the key and value results for reuse in subsequent steps. In the decode stage, the model generates the answer step by step, producing one token at a time. Unlike prefill, we only take the last token  in the sequence as our input because the previous tokens have been computed in the previous steps and only need to be called from the cache. Therefore, in the modules except for attention, we only need to calculate based on this single token. In attention, after computing the query, key, and value for the current input, we concatenate the saved key and value with the current value, allowing this token to obtain information from the previous sequence, as shown in the following formula.

\begin{align}
\{X^i_{key}, X^i_{value}\} \gets & Concat(\{X^{i-1}_{key}, X^{i-1}_{value}\}, \\
& \{token^i \cdot w^i_{key}, token^i \cdot w^i_{value}\}) \nonumber
\end{align}

\subsection{Memory Analysis}

\begin{figure}[h]
    \centering
    \includegraphics[width=0.5\linewidth]{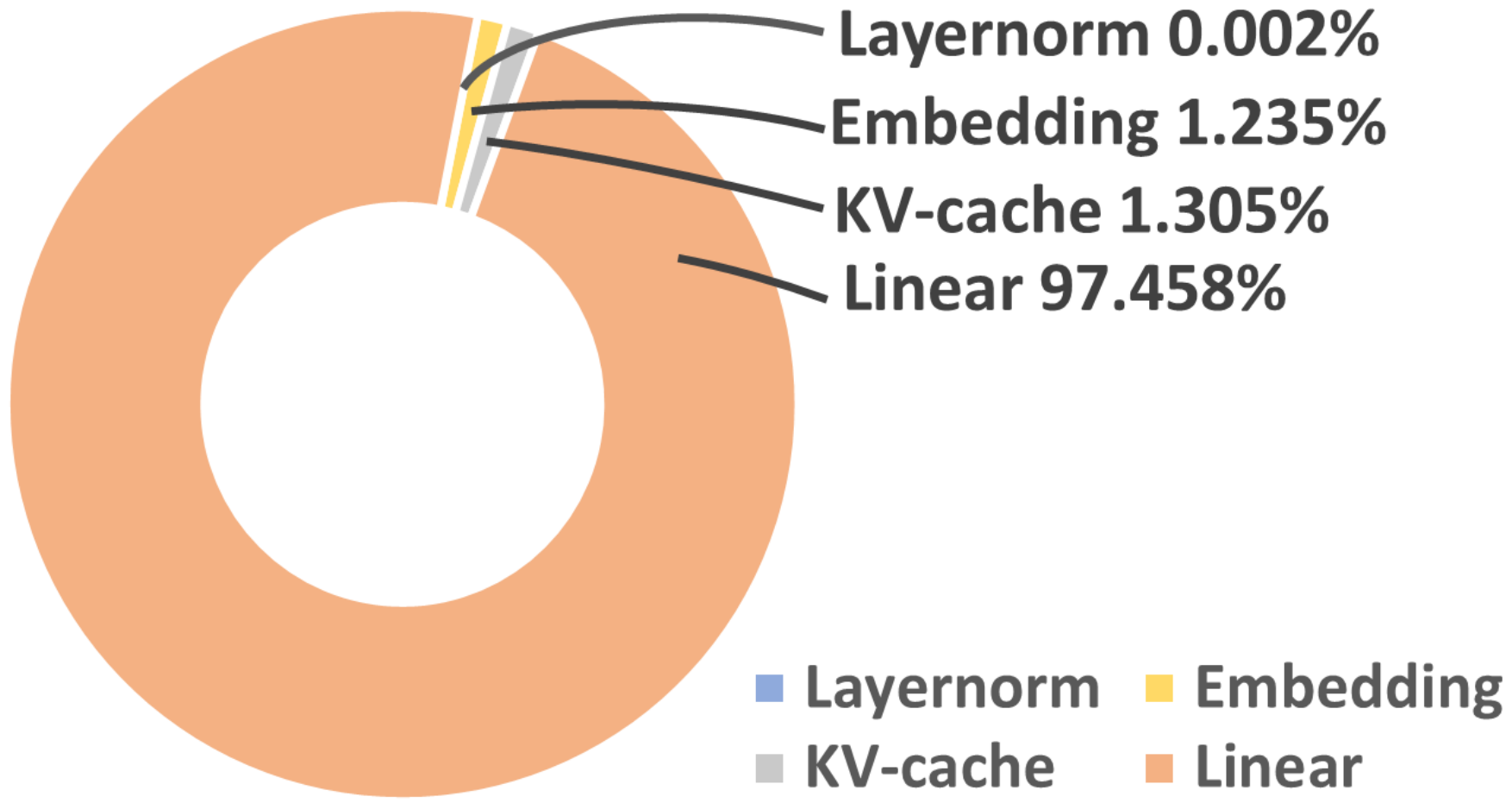}
    \caption{Memory usage in OPT-30B. Batch size is 1, and sequence length is 512.}
    \label{fig:weight_ratio}
\end{figure}

As depicted in Figure \ref{fig:weight_ratio}, we provide a insight into the memory utilization of OPT-30B. It becomes evident that the linear modules constitute a substantial majority, exceeding 97\% of the total memory consumption. In contrast, other components such as layernorm, embedding, and KV-cache make up only a minor fraction of the overall memory footprint. This observed proportionality extends to smaller model variants, reinforcing the significance of prioritizing offloading strategies for optimizing linear modules within the architecture.

\subsection{Offloading Bottleneck}
As illustrated in Figure \ref{fig:speed}, it is evident that the speed of I/O operations lags significantly behind that of the GPU, and even falls short of CPU performance by more than a tenfold margin. This highlights that the conventional approach, as suggested in prior works, which involves loading all parameters from the CPU to the GPU, is not the most efficient strategy. Instead, we can leverage the CPU to handle the computations as well. However, it is also impractical to have the CPU handle all workloads, as it would leave the GPU and I/O resources underutilized and wasted. Therefore, the optimal solution lies in a distribution of computation between the CPU and GPU, allowing them to work in parallel.

\section{Heterogeneous  Parallelism}
In this section, we discuss the design of our heterogeneous parallelism algorithm and computation distribution law. Section \ref{sec:para_strategy} focuses on exploring the most efficient parallelism strategy that aligns with our purpose. Furthermore, section \ref{sec:formula} investigates the guidelines for the distribution of computations to enable effective overlap among the CPU, GPU, and I/O under such parallelism strategy.

\subsection{Parallelism Strategy}\label{sec:para_strategy}

\begin{figure}
    \centering
    \begin{subfigure}{0.8\linewidth}
        \includegraphics[width=\linewidth]{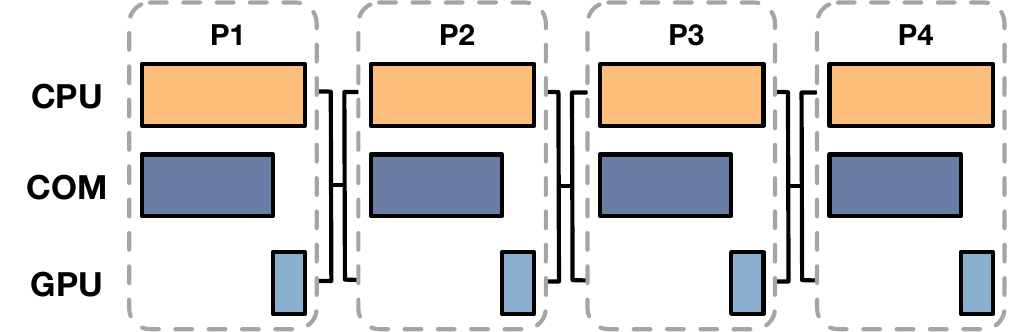}
        \caption{Heterogeneous tensor parallelism.}
        \label{fig:demo_tp}
    \end{subfigure}
    
    \vspace{5pt}
    
    \begin{subfigure}{0.8\linewidth}
        \includegraphics[width=\linewidth]{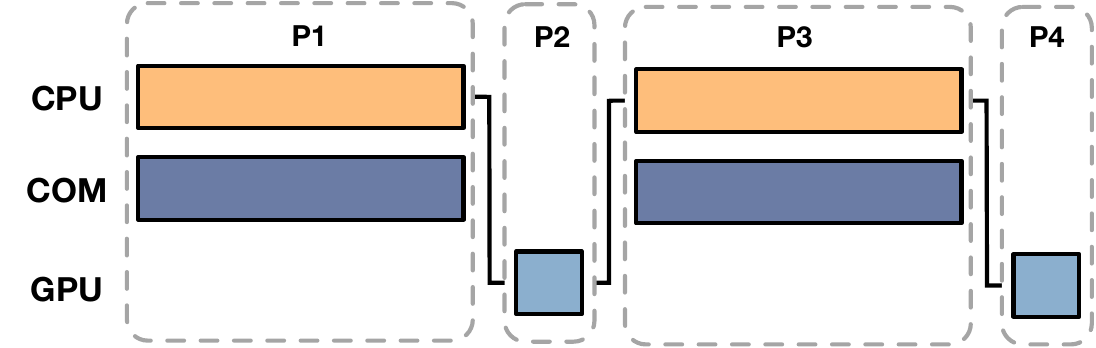}
        \caption{Heterogeneous pipeline parallelism.}
        \label{fig:demo_pp}
    \end{subfigure}
    \caption{A straightforward illustration of heterogeneous parallelism. COM denotes parameter communication, $P_i$ refers to the i-th part of the model and the black line represents data exchange.}
    \label{fig:demo}
\end{figure}

As elaborated in Section \ref{sec:related_parallel}, the field of deep learning encompasses three primary forms of parallelism: data parallelism \cite{zero}, tensor parallelism \cite{megatron}, and pipeline parallelism \cite{gpipe}. This subsection is dedicated to an in-depth exploration of the most effective parallelism strategy for GPU-CPU heterogeneous computing. A distinguishing factor in our task is the considerable disparity in processing speeds between the CPU and GPU components. The CPU computations and I/O operations exhibit speeds that are often tens or even hundreds of times slower than GPU computations. Consequently, it is imperative to meticulously address load balancing and the potential for communication overlap in order to optimize performance.

Data parallelism is evidently not a suitable choice for our scenario due to the small batch sizes our inputs, rendering them unsuitable for partitioning into mini-batches. Furthermore, the limited computational capacity of the CPU results in a substantial decrease in processing speed. On the other hand, both tensor parallelism and pipeline parallelism seems to offer promising solutions to this challenge. Considering the previously outlined distinctions, we present a simplified visualization of heterogeneous tensor and pipeline parallelism in Figure \ref{fig:demo}. In this visualization, we have omitted the consideration of communication overlap between layers for the sake of simplicity. It is apparent that GPU computation times are notably shorter due to their significantly superior efficiency in comparison to the CPU and communication. To minimize bottleneck communication time, we have strategically allocated a larger portion of the computation to the CPU to reduce communication volume.

To be specific, within the framework of tensor parallelism, the model is divided into two components, one designated for CPU processing and the other for GPU execution. The CPU is exclusively dedicated to computational tasks, and while CPU computation is underway, model parameters are conveyed to the GPU. The GPU, in turn, generates results once the communication process is completed. In this specific context, it is imperative to ensure:

\begin{equation}
    T_{CPU} = T_{GPU} + T_{COM}
\end{equation}

In order to fully harness the computational capabilities of the CPU and the communication, a method of achieving this involves partitioning the weight's dimensions. Furthermore, idle periods during GPU computations can be mitigated through prefetching.

In the context of pipeline parallelism, the model is segmented into distinct parts, and computations are executed sequentially on various devices. As evident from the illustration, in order to maximize utilization, it is crucial to ensure that the CPU computation time for the current part matches the communication duration for the subsequent stage, as denoted by:

\begin{equation}
    T^{P_i}_{CPU} = T^{P_{i+1}}_{COM} \label{equ:pp}
\end{equation}


In this scenario, we can discern the drawbacks of pipeline parallelism in comparison to tensor parallelism: 
1) CPU idleness during GPU computations is a notable drawback. Although the utilization of micro-batches is a common approach to alleviate this idle time, it proves challenging for our sparse input data, hindering overall efficiency.
2) A critical concern lies in the necessity to segregate different operators into distinct parts to satisfy Equation \ref{equ:pp} for pipeline parallelism. Achieving a perfect allocation is a difficult task, particularly for operators with varying computation times such as matrix multiplication, which inevitably leads to idle time. The option of stacking more layers into a single stage for better time balance is restricted by the GPU's finite parameter-handling capacity.
3) While pipeline parallelism generally incurs lower data exchange costs than tensor parallelism, this advantage becomes less important in the context of large language model generative inference, where activation sizes are relatively small. Consequently, for our specific task, tensor parallelism emerges as the optimal strategy.

\subsection{Computation Distribution}\label{sec:formula}
As mentioned earlier, parameter transmission has emerged as a significant bottleneck when conducting model inference on resource-limited devices. A practical strategy involves minimizing the data exchange between the GPU and memory and redistribute some parameter and computation to the CPU. To begin, we must decide how to distribute the workload between the GPU and CPU for any given module. Our primary aim is to balance the processing times of the GPU and CPU to optimize efficiency.

For the CPU, since the model parameters are already present in memory, the total CPU processing time is equivalent to the time required for parameter calculations on the CPU. On the other hand, for the GPU, parameter must be transferred from memory to GPU memory before computation can commence. Thus, the overall GPU processing time includes both data transfer time and GPU computation time.

Let's denote $T_{CPU}$ as the CPU processing time, $T_{GPU}$ as the GPU processing time, and $T_{COM}$ as the time for CPU-to-GPU data transfer. Our initial objective can be stated as Equation \ref{equ:euqal_time}. Next, let's assume that the total number of parameters is $W$, and the portion of parameters computed on the GPU is expressed as $\alpha$. The CPU's processing speed is denoted as $V_{CPU}$, and the GPU's processing speed is $V_{GPU}$. Additionally, the transmission speed is indicated as $V_{COM}$. We can elaborate on the equation as follows:

\begin{equation}
    \frac{(1-\alpha)W}{V_{CPU}} = \frac{\alpha W}{V_{GPU}} + \frac{\alpha W}{V_{COM}} \label{equ:euqal_time}
\end{equation}

After simplifying the expression, we arrive at the following formula:

\begin{align}
    \alpha & = \frac{V_{GPU}V_{COM}}{V_{CPU}V_{GPU}+V_{CPU}V_{COM}+V_{COM}V_{GPU}} \nonumber \\
    & = \frac{1}{\frac{V_{CPU}}{V_{COM}}+\frac{V_{CPU}}{V_{GPU}}+1}
\end{align}

We can disregard the GPU-to-CPU computation time ratio since it is usually negligible in the majority of cases, represented as:

\begin{equation}
    \alpha \approx \frac{V_{COM}}{V_{COM}+V_{CPU}}
\end{equation}

To further simplify, we can substitute $V$ with $1/T'$, where $T'$ signifies the overall duration of the entire operation, while the $T$ mentioned earlier pertains to the duration of the distributed computation. The ultimate formula can be expressed as:

\begin{equation}
    \alpha \approx \frac{{T'}_{CPU}}{{T'}_{CPU}+{T'}_{COM}} \label{equ:alpha}
\end{equation}

Hence, we derive a formula for the CPU-GPU heterogeneous computation ratio. Dividing the parameters based on this ratio enables us to establish an optimal proportion, maximizing theoretical efficiency. The significance of this formula is twofold:
1) It offers a straightforward and elegant description of distribution principles, applicable not only to Transformers but also to all forms of heterogeneous computation following this tensor parallelism approach.
2) Its practicality is noteworthy, as it requires only the measurement of execution times for these two operations in the respective models, making it exceptionally user-friendly for implementing such strategies.

\begin{figure*}[t]
    \centering
    \includegraphics[width=\textwidth ]{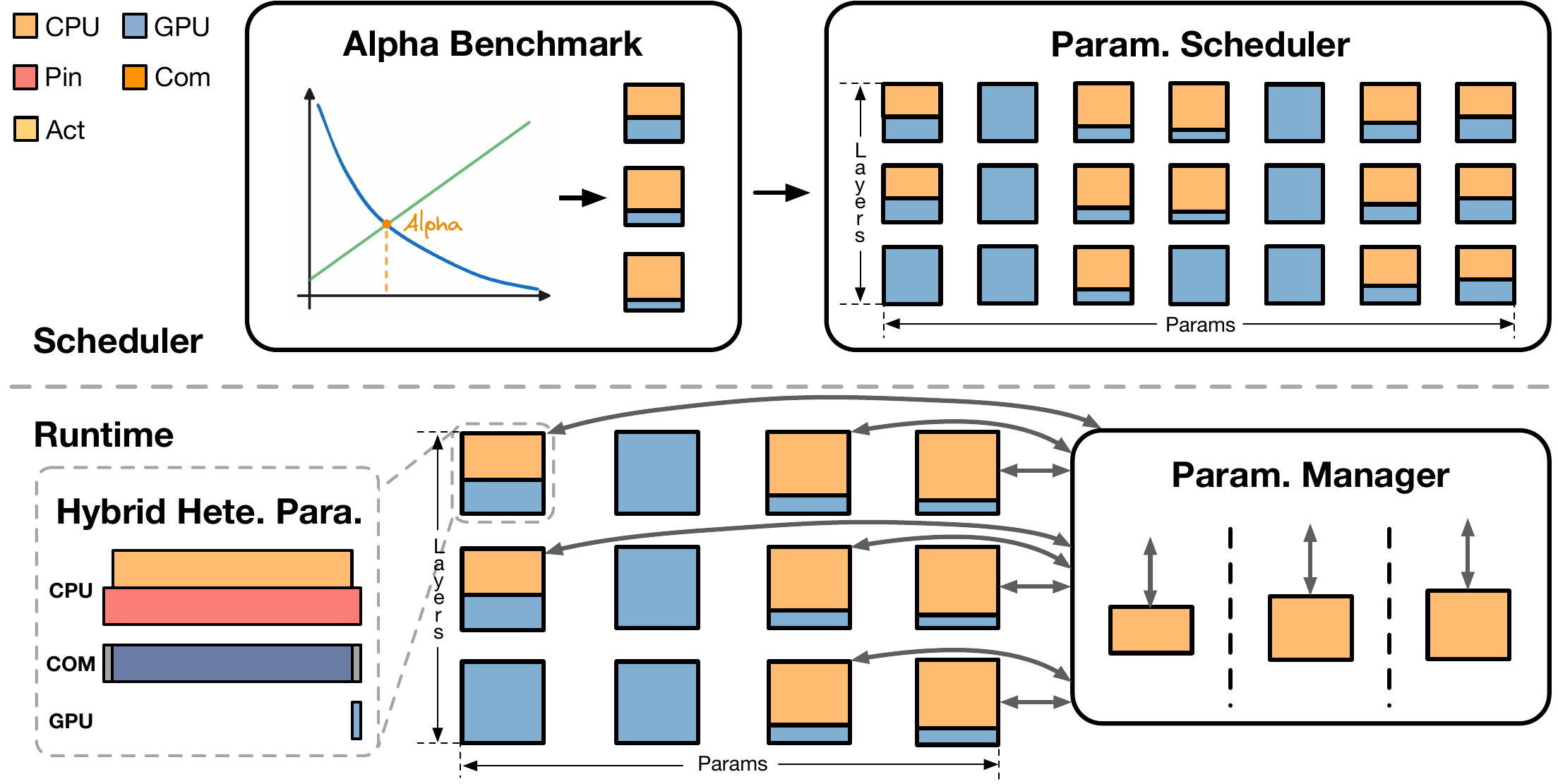}
    \caption{Overview of HeteGen. HeteGen has two main stages: scheduling and runtime. In the scheduling stage, it uses the alpha benchmark to distribute computation and decides on parameter policies based on our scheduler. In the runtime stage, it optimizes I/O and CPU utilization within heterogeneous modules using hybrid parallelism and manages asynchronous weights to minimize system impact.}
    \label{fig:hete}
\end{figure*}

\section{HeteGen}

\subsection{Overview}

To enable low-latency large language model inference on resource-constrained devices, we introduce HeteGen, a solution that harnesses parallel heterogeneous computing to enhance efficiency. HeteGen strategically exploits the capabilities of the CPU and I/O to mitigate the bottleneck I/O associated with offloading. This approach minimizes the need for parameter transfers between the GPU and memory, resulting in improved computational efficiency. To facilitate the implementation of this strategy, we outline a guideline for heterogeneous parallel computation in the previous section. In this section, we endeavor to apply this approach to the inference of large language models.

As illustrated in Figure \ref{fig:hete}, HeteGen consists of two main stages: the scheduler stage and the runtime stage. In the scheduler stage, HeteGen initially employs the alpha benchmark to determine the distribution ratio of computation for each module. Subsequently, the parameter scheduler assesses whether a heterogeneous policy should be applied to a module or place parameters solely on the GPU, based on our proposed value function and model configuration. In the runtime stage, within each heterogeneous module, our hybrid heterogeneous parallelism technique is employed to optimize their utilization of I/O and CPU, making use of asynchronous overlap. Additionally, a parameter manager oversees the management of these asynchronous weights to minimize their impact on the system.

\subsection{Hybrid Heterogeneous Parallelism}

\begin{figure}[t]
    \begin{subfigure}{\linewidth}
        \includegraphics[width=0.9\linewidth]{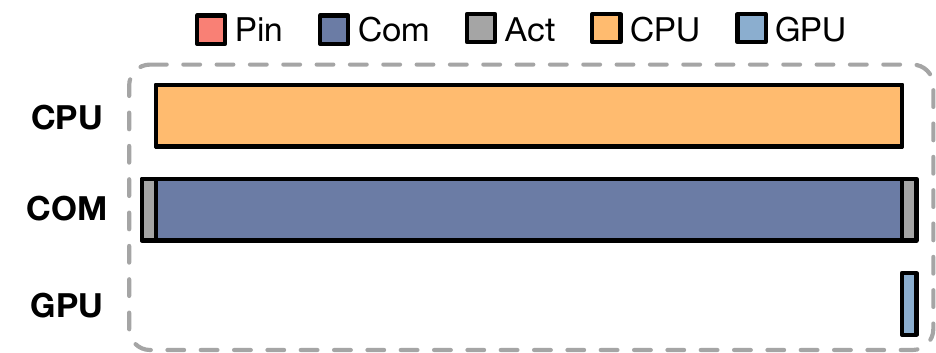}
        \caption{Naive heterogeneous parallelism.}
        \label{fig:hete_module1}
    \end{subfigure}
    \vspace{-5pt}

    \begin{subfigure}{\linewidth}
        \includegraphics[width=0.85\linewidth]{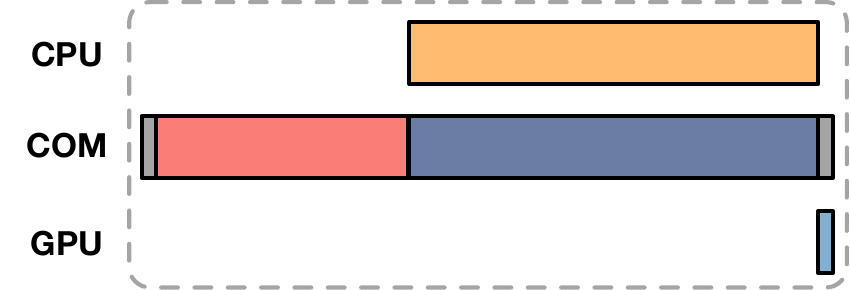} 
        \caption{Heterogeneous parallelism with pinned memory.}
        \label{fig:hete_module2}
    \end{subfigure}
    \vspace{-5pt}

    \begin{subfigure}{\linewidth}
        \includegraphics[width=0.7\linewidth]{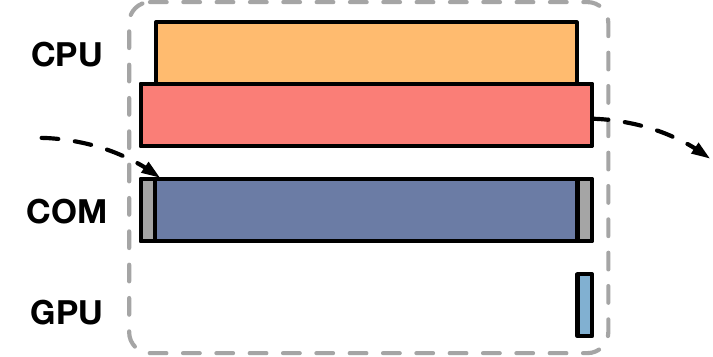} 
        \caption{Hybrid heterogeneous parallelism.}
        \label{fig:hete_module3}
    \end{subfigure}
    \caption{Demonstration of different heterogeneous parallelism strategies.}
    \label{fig:hete_module}
\end{figure}

This section delves into the implementation details of HeteGen's hybrid heterogeneous parallelism approach. As illustrated in Figure \ref{fig:weight_ratio}, we observe that the majority of memory is consumed by the linear in LLMs. As a result, we focus our attention on applying heterogeneous parallelism to linear modules, while retaining all other modules on the GPU. To achieve the best efficiency with our hybrid heterogeneous parallelism, the key to reducing latency is as follows: 1) Ensure the system bottlenecks, notably CPU computation and CPU-GPU communication, are fully utilized. 2) Maximize the overlap between communication and CPU computation. Equation \ref{equ:alpha} provides insight into the optimal distribution of computation between the CPU and GPU for optimal utilization. With $\alpha$ already determined, our focus shifts to optimizing the overlap within a single module.

As depicted in Figure \ref{fig:demo_tp}, it provides a naive outline for the design of heterogeneous parallelism. Following this framework, we can implement the most straightforward form of asynchronous overlap, as illustrated in Figure \ref{fig:hete_module1}. To clarify, the process first involves transmitting activation from the GPU to the CPU, then commencing CPU computation, and asynchronously transferring GPU weights to the CPU. Once the CPU computations are complete, the outputs are sent back to the GPU for final processing. However, this basic form of heterogeneous parallelism is not the most efficient since it fails to fully utilize the I/O bandwidth between the CPU and GPU. For better bandwidth, some prior works \cite{flexgen} employ pinned memory to expedite weight transfer, as illustrated in Figure \ref{fig:hete_module2}. This approach involves pinning the relevant CPU memory first, as pinned memory offers higher transfer speeds. But the pinning memory blocks both communication and CPU computation, leading to worse performance.

HeteGen has discovered that this approach is not the most efficient I/O overlap strategy, prompting us to propose a novel level of parallelism for memory operations, as demonstrated in Figure \ref{fig:hete_module3}. The diagram illustrates our hybrid heterogeneous parallelism. To elaborate, building upon the parallelism of CPU computation and CPU-GPU communication, we introduce a new form of parallelism dedicated to communication. This communication parallelism is divided into two components: pin memory and memory transfer, both of which operate concurrently using multithreading. However, this parallelism introduces a higher level of asynchrony since it necessitates the prior pinning of the current weight. We prepare the current pinned weight for transfer and proactively pin the next weight in the upcoming layer. This enables simultaneous execution of CPU computation, weight pinning, and weight transfer. Given the division of communication into pinning and transferring, our heterogeneous formulas in Equation \ref{equ:euqal_time} and Equation \ref{equ:alpha} require updating as follows:

\begin{align}
    T_{CPU} &= T_{GPU} + T_{COM}\nonumber \\
    &= T_{GPU} + max({T}_{PIN}, {T}_{TRANS}) \\
    \alpha &\approx \frac{{T}_{CPU}'}{{T}_{CPU}'+max({T}_{PIN}', {T}_{TRANS}')} \label{equ:async_alpha}
\end{align}

\subsection{Asynchronous Parameter Manager}

\begin{figure}[t]
    \centering
    \includegraphics[width=\linewidth]{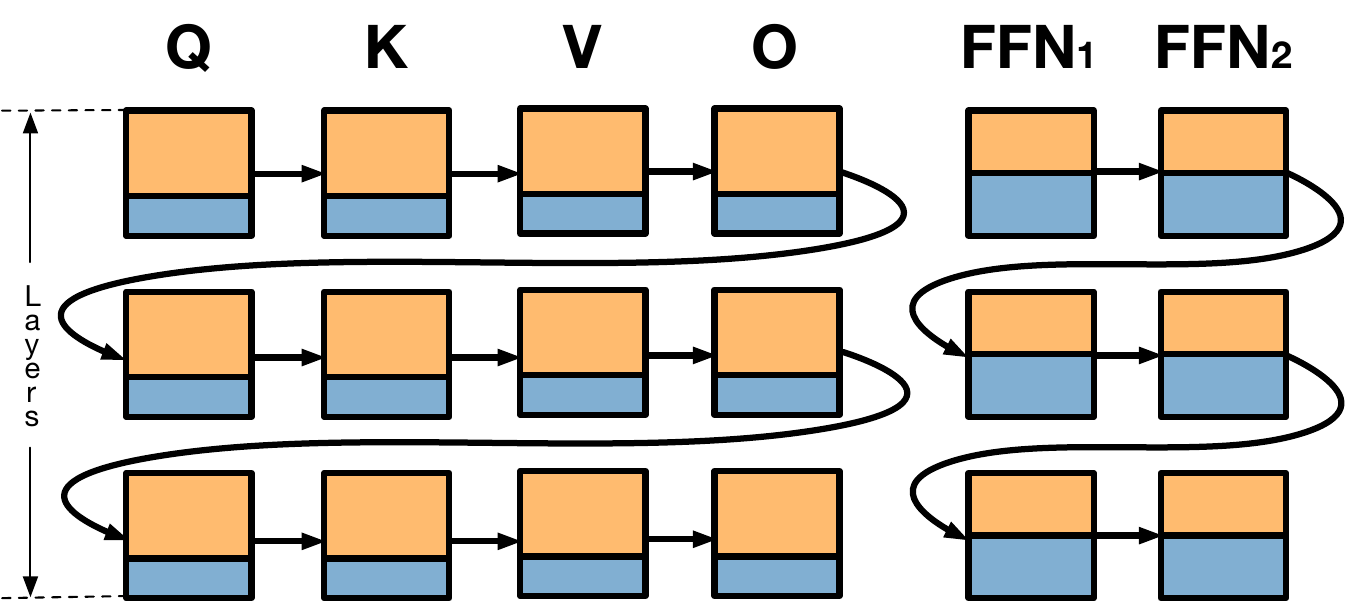}
    \caption{Illustration of asynchronous heterogeneous scheduler. The arrows refer to the pinned memory pass.}
    \label{fig:schedule}
\end{figure}

Given that the hybrid heterogeneous parallelism requires the asynchronous pinning of parameters for each heterogeneous module before computation, HeteGen introduces an asynchronous parameter manager. This manager is specifically designed to efficiently manage the temporary pinned parameter memory of our hybrid heterogeneous parallelism for each module.

The management of this process serves two primary objectives: maintaining asynchrony without affecting operational speed and minimizing memory and I/O costs associated with cached memory. Since the pinning time is contingent on parameter size, we categorize all heterogeneous parameters into two groups based on their size: linear in attention and linear in mlp. As depicted in Figure \ref{fig:schedule}, within each group, for every layer, the preceding heterogeneous module prepares the pinned weights for the subsequent parameters while acquiring its own pinned weights from the previous parameter. In the event that it is the last module within a layer, it proceeds to process the weights for the first parameter in the following layer. This approach offers two distinct advantages. Firstly, since the weight sizes within a group are uniform, the pinning times are nearly identical, preventing any bubble time during the process. Secondly, the approach restricts the system memory overhead to just a maximum of one pinned parameter per group.

In the scheduler, our initial step involves determining the alpha ratios for different modules. Based on these results, we make informed decisions about whether each module should undergo Hete Parallelism. If the answer is affirmative, we assign the corresponding $\alpha$ value to the module. Within each module, we execute computations utilizing the heterogeneous parallelism method described in the previous section. This approach ensures that, at any given moment, the GPU's memory is occupied by parameters equivalent to only $\alpha$ times the parameter size. As a result, our GPU memory requirements are significantly reduced without any adverse impact on efficiency. Building on this foundation, we further enhance the determination of alpha values and parameter scheduling, taking into account the specific characteristics of the computations.


\subsection{Alpha Benchmark}
As per our heterogeneous formula introduced in Equation \ref{equ:async_alpha}, the determination of the final weight distribution ratio $\alpha$ requires measurements of CPU computation speed and CPU-to-GPU transfer speed. Nevertheless, CPU computation and I/O speed do not exhibit a direct proportionality to the number of parameters, which is governed by $\alpha$. Benchmark results may also be influenced by system conditions, rendering them less stable and potentially divergent from their actual values.

To obtain more precise results, we employ a refined approach. Building upon the prior benchmark value $\alpha$, we adjust its value within a small range of $[\alpha + \gamma, \alpha - \gamma]$ in steps of $\lambda$ for minimum cost. This adjustment is followed by testing the times $T_{CPU}'$ and $max({T}_{PIN}', {T'}_{TRANS})$ as mentioned in Equation \ref{equ:async_alpha}. We then utilize polynomial formulas to model their speeds corresponding to different $\alpha$ values. This enables the calculation of the $\alpha$ value at which both speeds are equal, as depicted in Figure \ref{fig:hete}. This value is denoted as:

\begin{align}
    F_{CPU}(\overline{\alpha}) = F_{COM}(\overline{\alpha})
\end{align}

where $F$ refers to the fitted function.

\begin{align}
    &\overline{T}_{CPU} = F_{CPU}(\overline{\alpha}) \\
    \overline{T}_{COM} = max&(\overline{T}_{PIN}, \overline{T}_{TRANS}) = F_{COM}(\overline{\alpha})
\end{align}

\subsection{Heterogeneous Module Scheduler}

\begin{figure}[t]
    \centering
    \includegraphics[width=0.9\linewidth]{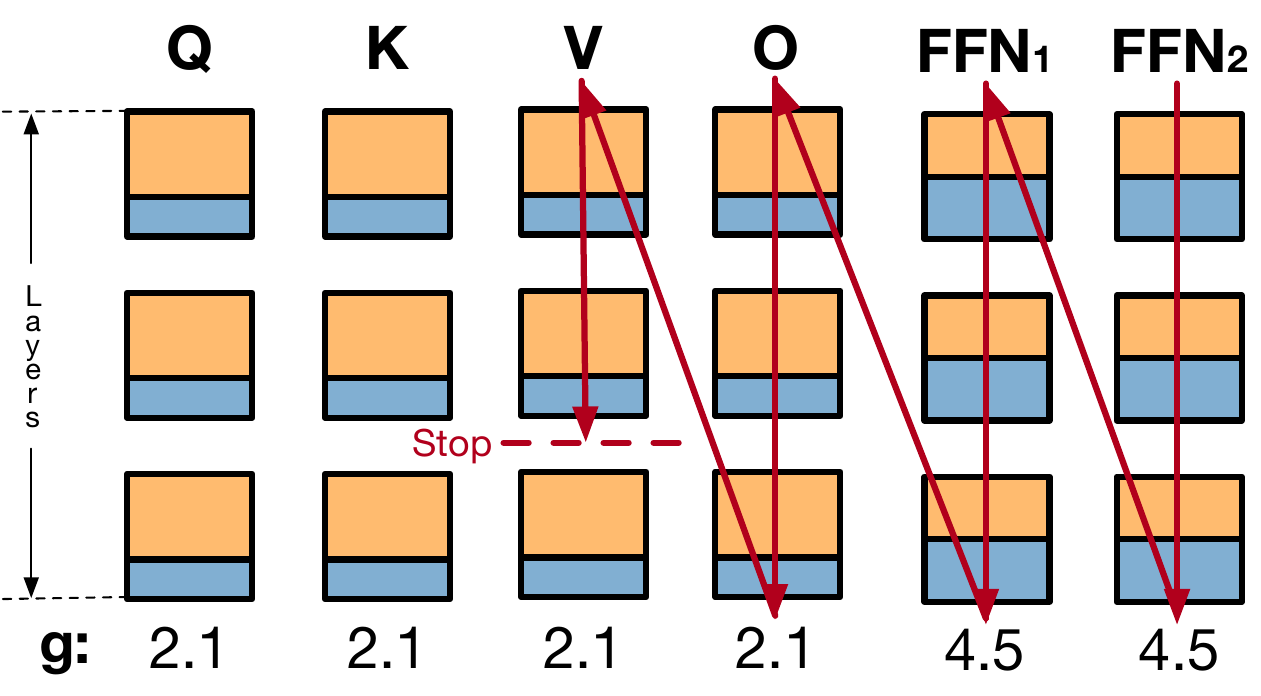}
    \caption{Illustration of heterogeneous module scheduler. The arrows refer to the order to place parameters to GPU.}
    \label{fig:param_place}
\end{figure}

Our scheduling approach effectively addresses the issue of high GPU memory demand by minimizing the parameters on the GPU. This ensures efficient memory utilization and low GPU memory usage. However, in scenarios with ample memory space available, our approach falls short of fully exploiting the GPU's resources.

To overcome this limitation and better utilize GPU memory, we introduce the Heterogeneous Module Scheduler. This technique dynamically transfers a portion of parameters from the CPU to the GPU when sufficient memory is available, reducing the original communication costs associated with these parameters. By allocating parameters between the CPU and GPU on the fly, the parameter schedule optimizes resource allocation, leveraging GPU capabilities while minimizing communication overhead. This enhancement positively impacts system performance and GPU utilization for our scheduling task.

The order in which parameters are allocated to the GPU is crucial, as the benefits of moving different weights to the GPU can vary. To assess the advantages of placing a module on the GPU, we can quantify it by considering the ratio of the time saved to the GPU memory consumption. Based on our previous formula, the saved time equals the module computation time, which is equivalent to our benchmarked CPU time $\overline{T}_{CPU}$. Thus, this metric can be denoted as:

\begin{equation}
    g = \frac{\overline{T}_{CPU}}{Mem}
\end{equation}

In particular, as illustrated in Figure \ref{fig:hete}, we can establish the ranking of each parameter by comparing their schedule gain ($g$). We then proceed to migrate the weight with the highest $g$ to the GPU for each layer until the memory limit is reached.



\section{Experiments}
\begin{figure*}[t]
    \centering
    \includegraphics[width=\textwidth ]{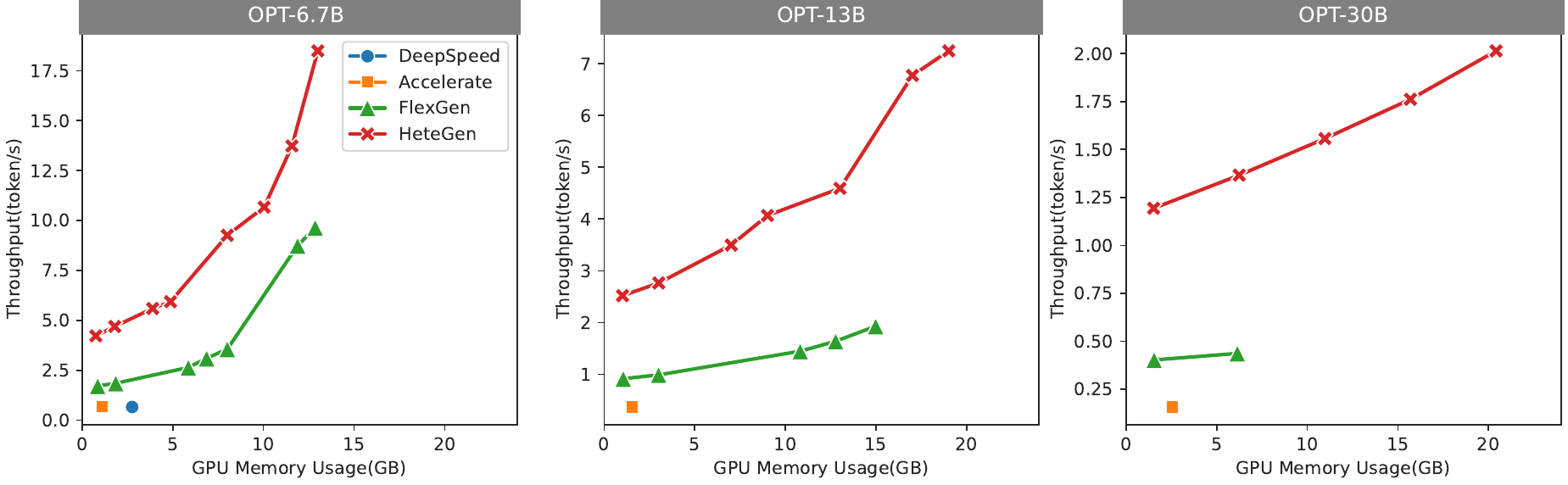}
    \caption{Throughput comparison under different memory constraints for OPT models from 6.7B to 30B with batch size=1, prefill length=512, decode length=64. DeepSpeed Inference and Accelerate cannot set setting the parameter offloading ratio. And DeepSpeed Inference is out of memory for OPT-13B and OPT-30B.}
    \label{fig:compare}
\end{figure*}

\begin{table}[h]
\centering
\caption{Hardware for Evaluation.}
\vspace{5pt}
\begin{tabular}{ccc}
\toprule
GPU & CPU & PCIE \\
\midrule
NVIDIA A10 & Intel Xeon @ 2.30GHz & 30GB/s\\
\bottomrule
\end{tabular}\label{tab:hardware}
\end{table}

\textbf{Hardware.} We run experiments on NVIDIA A10(24GB) GPU. GPU and CPU communicate via PCIE. The hardware specifications are listed in Table \ref{tab:hardware}. Our methods and implementations do not depend on specific hardware architectures. Some architecture with better GPU and CPU communication speed could be more friendly to our method.

\textbf{Model.} We use OPT models \cite{opt_2022} ranging in size from 6.7B to 30B to test HeteGen and other baseline methods. Although we do not evaluate other models, the offloading in HeteGen can be applied to other large language models, e.g., LLaMA \cite{llama_2023}, GPT-3 \cite{gpt3}, and BLOOM \cite{bloom_2023} because they all share a similar transformer structure.

\textbf{Workload} We tested with prefill lengths 512 with batch size 1, and compare the throughput of generating tokens under different scenarios where the generated length is 64 to evaluate the effectiveness of different methods on a single GPU and limited CPU memory. And we limit CPU cores using to be at most 16.

\textbf{Baseline.} We used DeepSpeed Inference \cite{deepspeed_inference}, HuggingFace Accelerate \cite{accelerate}, and FlexGen \cite{flexgen} as baselines, all of which support offloading the weight parameters of the model. Deepspeed Inference and HuggingFace Accelerate implements naive offloading strategy and FlexGen utilize CPU computation for attention and batch schedule to improve efficiency for large batch. In our experiments, since HeteGen focuses on reducing the latency of running large models, we did not offload data to disk.

\textbf{Implementation.} HeteGen is implemented based on PyTorch \cite{pytorch_2019} and uses CUDA streams and CPU threads to enable overlap.


\subsection{End-to-End Performance}

We conducted an evaluation of the throughput for each method, considering various GPU memory utilization scenarios by adjusting the offload parameter ratios. We compared their throughput when operating under similar memory consumption conditions. The performance comparison results are presented in Figure \ref{fig:compare}, revealing that our approach consistently outperforms all other methods while achieving lower memory utilization for the same level of speed. HeteGen also offers a wider range of dynamic offload adjustments.

Accelerate and DeepSpeed Inference lack the capability for automatic dynamic adjustment of GPU and CPU weight proportions based on offload ratios, so they are presented as dots in the figure. DeepSpeed Inference experiences out-of-memory (OOM) issues when handling OPT-13B and OPT-30B models.
Both FlexGen and HeteGen demonstrate support for dynamic workload adjustments between the GPU and CPU. However, under equivalent GPU memory consumption conditions, HeteGen consistently showcases superior throughput compared to FlexGen. HeteGen demonstrates an increase of up to 317\% in the case of OPT-30B models. And the dynamic adjustment range of HeteGen significantly surpasses that of FlexGen, owing to our more advanced scheduling strategy. Specifically, for OPT-30B models, HeteGen effectively adapts to GPU memory usage ranging from 6.5\% to 88.7\%, whereas FlexGen is limited to supporting scenarios ranging from just 6.5\% to 26.5\%, and it encounters OOM issues beyond this threshold.

Moreover, HeteGen underscores the fact that, irrespective of the different memory constraints, as long as your hardware cannot accommodate all model parameters solely on the GPU, it has the potential to enhance inference latency by a large margin, proving its effectiveness and genericity.

\subsection{Runtime Breakdown}

\begin{table}[h]
\centering
\caption{HeteGen runtime break down for OPT-13B in a heterogeneous linear module.}
\vspace{5pt}
\begin{tabular}{ccccc}
\toprule
All & CPU & I/O & Pin & GPU  \\
\midrule
100\% & 97.8\% & 96.9\% & 72.4\% & 0.1\% \\
\bottomrule
\end{tabular}\label{tab:runtime_breakdown}
\end{table}

As shown in Table \ref{tab:runtime_breakdown}, we illustrate the breakdown time of HeteGen for OPT-13B within a heterogeneous linear module, to be specific, the first linear in MLP. It is evident that the CPU and I/O are nearly fully utilized, with the exception of kernel launch time. Additionally, the time taken for pin memory operations is shorter than that for I/O, so it is easy to be overlapped. This observation suggests that the CPU, I/O, and pin operations are effectively overlapped, substantiating the efficiency of HeteGen.

\subsection{Ablation Study}

\begin{table}[h]
\centering
\caption{Ablation study for HeteGen.}
\vspace{5pt}
\begin{tabular}{lc}
\toprule
Methods            & Performance \\
\midrule
All                & 100\%       \\
no hybrid heterogeneous parallelism & 77.7\% \\
no asynchronous parameter manager & 94.9\% \\
no alpha benchmark  & 92.8\% \\
no heterogeneous module scheduler & 32.1\% \\
\bottomrule
\end{tabular}\label{tab:albation}
\end{table}



In ablation study, we evaluated the performance of our optimizations, including heterogeneous parallelism, asynchronous parameter manager, alpha benchmark, and heterogeneous module scheduler. As shown in Table \ref{tab:albation}, we evaluate the effectiveness of these optimizations. We can see in the table that alpha benchmark and async parallel contributes a lot to the performance. And buffer I/O indeed reduce I/O time with 2.1\%.

\section{Related Work}
\subsection{Memory Optimization}
Numerous studies are currently investigating ways to improve inference efficiency on devices with limited resources. In this paper, we will examine how existing research has accelerated large language model inference and reduced resource consumption from three perspectives: low-level optimization, model optimization, and offload optimization.

\subsubsection{Offloading}
Offloading is currently one of the most effective methods for reducing GPU memory usage. By temporarily storing unused parameters in memory or disk, offloading greatly reduces the demand for GPU memory. Techniques such as DeepSpeed Inference and Hugging Face Accelerate have utilized offloading to reduce memory usage during inference. However, this approach introduces data transfer as a bottleneck in inference, as transfer speeds are generally slower than computation speeds by one to two orders of magnitude. To address this issue, FlexGen \cite{flexgen} effectively increased system throughput by designing an offload strategy with larger batch sizes and using heterogeneous methods to compute attention. However, the latency of offload systems has yet to be well resolved.

\subsubsection{Model Compression}
Model optimization includes techniques such as quantization, pruning, and distillation. Quantization \cite{smoothquant, gptq} compresses the number of bits used to represent the model parameters and activations to reduce their memory usage. Pruning \cite{liu2017learning} reduces memory usage by removing redundant parameters, connections, or layers from the model. Distillation \cite{hinton2015distilling}uses a smaller model to learn the knowledge contained in a larger model. These methods can effectively reduce memory usage. However, as previously mentioned, they can only reduce a portion of the parameters and memory usage, which may still be insufficient for low-resource devices.

\subsubsection{Low-level Optimization}
Low-level optimization techniques, such as module scheduling, memory management \cite{fang2021turbotransformers}, and kernel optimization \cite{dao2022flashattention}, are crucial factors for improving inference efficiency. Many existing inference methods have significantly enhanced their speed and efficiency by utilizing these low-level optimization techniques, such as FasterTransformer, TurboTransformer \cite{fang2021turbotransformers}, Energon-AI \cite{du2022energonai}, DeepSpeed Inference \cite{aminabadi2022deepspeed}, and Hugging Face Accelerate \cite{accelerate}. Although low-level techniques can significantly increase inference speed, they are unable to effectively reduce GPU memory consumption.

\subsection{Parallelism}\label{sec:related_parallel}
There are three common parallelism for deep learning: data parallelism, tensor parallelism and pipeline parallelism. Data parallelism splits data into batches, processes them concurrently on multiple devices, and aggregates gradients for parameter updates. Tensor parallelism involves splitting a deep learning model's weight and activation across multiple devices, enabling the training of large, memory-intensive models. It partitions the model's layers or components across devices. Pipeline parallelism divides a model into stages, each processed by different devices. Data flows through these stages sequentially, improving efficiency for complex models.

\section{Conclusion}
In summary, our work introduces HeteGen, a heterogeneous parallel inference system aimed at significantly reducing latency in the context of Large Language Models (LLMs) on resource-constrained devices. It achieves this by harnessing the combined capabilities of both CPUs and I/O resources, making it a versatile solution, even when a system struggles to accommodate all parameters, including advanced GPUs. This marks a significant step forward in addressing the challenges posed by the ever-expanding models in the field of AI, with the ultimate objective of enhancing their accessibility and fostering democratization. Furthermore, further research and development in this direction hold the promise of even more remarkable advancements in facilitating efficient AI inference across diverse computing environments, potentially entailing the integration of quantization techniques and methods to reduce memory costs.

\section*{Acknowledgements}
Yang You's research group is being sponsored by NUS startup grant (Presidential Young Professorship), Singapore MOE Tier-1 grant, ByteDance grant, ARCTIC grant, SMI grant (WBS number: A-8001104-00-00),  Alibaba grant, and Google grant for TPU usage.





\bibliography{example_paper}
\bibliographystyle{mlsys2024}



\end{document}